# Direct Base-to-Base Transitions in ssDNA Revealed by Tip-Enhanced Raman Scattering

Xiu-Mei Lin [†,‡], Tanja Deckert-Gaudig[†]*, Prabha Singh[◊], Michael Siegmann[◊], Stephan Kupfer[◊], Zhenglong Zhang[†], Stefanie Gräfe[◊], Volker Deckert[†,◊]*

[†]IPHT - Leibniz Institute of Photonic Technology, Albert-Einstein-Str. 9, D-07745 Jena, Germany

[◊]Institute of Physical Chemistry and Abbe Center of Photonics, Friedrich-Schiller-Universität Jena, Helmholtzweg 4, D-07743 Jena, Germany

[‡]present address: HIU-Helmholtz-Institute Ulm, Helmholtzstr. 11, D-89081 Ulm

**Abstract**

In the present contribution, specifically designed single-stranded DNA (ssDNA) sequences composed of adenine and cytosine were used as nanometric rulers to target the maximum achievable spatial resolution of tip-enhanced Raman spectroscopy (TERS) under ambient conditions. By stepping along a strand with a TERS tip, the obtained spectra allowed for a clear spectral discrimination including conformational information of the nucleobases, and even sharp adenine-cytosine transitions were detected repeatedly with a spatial resolution below 1 nm.

**Introduction**

Recently, nanoscale analysis has become more and more important for achieving a deeper understanding of chemical processes and molecular mechanisms in biochemistry. In particular, genome and cancer related research could benefit from the identification of mismatches in DNA or the localization of methylated nucleobases.[1] Complementary contributions in these fields can be provided by tip-enhanced Raman spectroscopy (TERS), which is a combination of Raman spectroscopy and scanning probe microscopy (i.e., atomic force microscopy - AFM or scanning tunneling microscopy - STM).[2-5] This technique facilitates label-free morphological and chemical characterizations of a sample in a single experiment with nanoscale resolution. Replacing standard AFM tips with specifically metalized (Ag or Au) probes for TERS, the intrinsically low sensitivity of Raman spectroscopy can be overcome, and even single molecule levels become accessible as demonstrated for strong Raman scatterers under specific conditions.[2, 6] It is noteworthy that the high spatial accuracy of AFM/STM as well as the high spectral specificity of Raman spectroscopy is maintained. The basic concept of the signal enhancing effect is essentially the same as in surface-enhanced Raman scattering (SERS), where localized surface plasmon resonances in rough metal surfaces are excited under illumination. Those plasmons affect vibrational modes of molecules in the closest vicinity to the metal surface, yielding drastically enhanced Raman signals. This enhancement is composed of two mechanisms: A dominating electromagnetic part and a chemical part, where the latter results primarily from metal-molecule charge-transfer interactions.[7-11] In TERS, the SERS substrate is reduced to a single Ag/Au nanoparticle or even an edge that acts as a scanning probe and a Raman signal enhancing device simultaneously.

Lately, it has been demonstrated that TERS has the potential to reach a spatial resolution below 10 nm for protein structures[12], carbon nanotubes[13, 14] and DNA[5, 15] at ambient conditions. Under certain experimental conditions (i.e., gap-mode, single molecule, high vacuum, and low temperature), resolutions even down to 0.5 nm could be realized.[16-18] Those extreme values seem to surpass the expected resolution of 5-10 nm predicted by the classical electromagnetic theory considering particle sizes of tens of nanometers.[19, 20] In general, it is assumed that the spatial resolution of TERS depends on many parameters, including the nanoparticle at the tip apex (size, shape, metal), the nature of the highly confined electromagnetic field, the tip-sample distance and even chemical interactions.[21-24] Generally, it has been assumed that the size of the metallic tip predominantly determines the lateral resolution, indicating that a (sub-) nanometer resolution is out of reach. However, recent ultra-high vacuum low temperature (UHV-LT) TERS results have convincingly demonstrated that this conventional concept needs to be reconsidered.[16, 17] In these studies, the authors obtained a spatial resolution which enabled the distinction of molecular sites of tetraphenylporphyrins. Scanning electron microscopy analyses have shown that the diameters of AFM and STM TERS tips are at least 10 nm and in many cases are much bigger, but evidently, the spatial resolution is not necessarily coupled to the apex size. Current efforts to explain the gap between experimental observations and theory treat the entire tip-sample system or consider atomic-size sharp edges in full quantum mechanical approaches.[21, 25, 26] Another important issue in TERS is the depth resolution. Theoretical predictions are in the range of a few nanometers[27] and are in agreement with experimental estimates, e.g., no detection of the dominating β-sheet core of insulin fibrils[28], the identification of "buried" cytochrome c in mitochondria[29] and the absence of dominant nucleobase signals in double-stranded DNA[30].

In this contribution, single-stranded DNA (ssDNA) with predefined sequences was chosen to target the maximum spatial resolution in TERS under ambient conditions. ssDNA is an ideal system for this endeavor because its linearity reduces the spectral changes in TERS to a single dimension. Double-stranded DNA is impractical here because the phosphate residues are facing outside, whereas the nucleobases are protected from direct tip access inside. In DNA, the nucleobases adenine (A), cytosine (C), thymine and guanine are bound to deoxyribose, which in turn is bound to phosphate groups forming the backbone. Nucleic acids have already been the subject of numerous SERS studies, see e.g., Refs. 31-36. An important conclusion of these investigations is that the DNA-substrate configuration has a clear impact on the spectra. On SERS substrates, DNA usually binds at an energetic favored geometry to the metal, which is influenced by the material, pH, etc. Hence, spectra appear quite different and show band shifts and intensities depending on the coordination site and metal-sample interactions. Nevertheless, the characteristic ring breathing mode, in particular, enables an unambiguous differentiation of the nucleobases. Consequently, for TERS on DNA, a defined strand adsorption on the substrate is mandatory to achieve consistent signal detection. To obtain reproducible DNA spectra with TERS, strands have to be immobilized in a more or less stretched geometry. Early experiments on such stretched homopolymer strands of adenine and cytosine validated that consistent spectra can indeed be obtained.[37, 38] For randomly oriented DNA on the other hand, spectra can vary significantly.[5, 30, 34, 39, 40] To avoid intra-strand base pairing, ssDNA with a known sequence containing only adenine and cytosine was chosen. The heteroaromatic rings allow both DNA bases to be good Raman scatterers, which can be well distinguished. In particular, the characteristic ring breathing mode enables straightforward identification and discrimination. Adenine has been the subject of numerous experimental and theoretical SERS works with the focus on its coordination chemistry and resulting spectral changes, see e.g., Refs. 34, 41-43. It has also been studied frequently by TERS, [34, 44-46] where especially data on a single adenine crystal have provided valuable insights into the effect of a changing tip-molecule geometry on the probing of different molecular sites.[44] Cytosine has been characterized to the same extent, see e.g., Refs. 34, 46-49. Interestingly, the previous research indicates that a comparison between TERS and SERS and conventional Raman is not straightforward. A specific point to consider in TERS is that spectral information is acquired from only a few molecules, and therefore, even for non-specifically designed single molecule experiments, averaging effects are absent or by far not as dominant as in "bulk" SERS or Raman spectra.[16, 50, 51] Furthermore, the tip-sample distance also influences signal intensity and band position.[52] Consequently, TERS measurements on homopolymer strands of Poly(A)[38] and Poly(C)[37] can be considered as single (large) molecule experiments, and small variations in the tip-strand orientation and distance must induce spectral changes, i.e., band shifts and intensities. This behavior is similar to other single molecule experiments, such as fluorescence or SERS, where intensity variations and spectral shifts originate from a specific tip-specimen geometry, energy and excited state lifetime rather than from diffusion and electromagnetic field gradients.[50]

**Experimental**

Single-stranded DNA with a sequence of $(A_{10}C_{15})_8$ and $(ACAC_3AC_5AC_{10}AC)_8$ (Integrated DNA Technologies, Belgium) were used without further purification. The DNA was dissolved in 10 mM HEPES (4-(2-hydroxyethyl)-1-piperazineethanesulfonic acid) to maintain a physiological pH value, and 10 mM magnesium chloride solution was used to fix the phosphate site onto the mica surface. The DNA solution (c = 10 ng /μL) was heated to 90 °C for $(A_{10}C_{15})_8$ and 80 °C for $(ACAC_3AC_5AC_{10}AC)_8$ for 10 min, followed by rapid cooling in an ice bath for 10 min to stretch the DNA single strand.[37] 2 μL of the thermally treated DNA solution was then dropped onto a freshly cleaved mica sheet and incubated for 5 min (($A_{10}C_{15})_8$) and 10 min (($ACAC_3AC_5AC_{10}AC)_8$). Finally, samples were washed twice with distilled water to remove the buffer and then dried. TERS spectra were recorded along the strand with a step-size of 0.3 nm at $\lambda_{exc}$ = 532 nm (P = 750 μW, $t_{acq}$ = 5 s) using a setup composed of an AFM (Nanowizard 2, JPK, Germany) mounted on top of an inverted microscope (Olympus IX 71, Japan). The microscope was equipped with a 60x oil immersion objective (N.A. 1.45) and connected to a Raman module (S&I, Germany). The root-mean-square of the lateral repeatability of the instrument was 0.3 nm according to the specifications of the manufacturer. A control experiment yielded a drift of 30 nm/hour after thermal equilibration of the experiment. Spectra were analyzed with Igor PRO 6.35 (Wavemetrics, USA). All spectra shown in the manuscript were raw data and analyzed without further baseline correction or smoothing with non-linear band fitting.

**Results**

For the first TERS experiments, ssDNA with the distinct sequence $(A_{10}C_{15})_8$ was used to investigate whether a base-to-base transition can be detected by spectral differences when pacing along the strand. Considering the strong coupling of topography and light signal, a measurement along the main axis of the strand is more reasonable for estimating the spatial resolution than measuring in the perpendicular directions. This way, topographic coupling, i.e., a major contribution to "false assignment" of the resolution, is avoided.[53] Stepping with TERS across the convex/curved strand, in which the width in the topography always appears larger than it actually is, results in a much lower apparent resolution as was shown previously.[38] Therefore, the one-dimensional main axis of a single-stranded DNA is ideal for determining the resolution because the sample can be considered as flat along the strand. The inter-phosphate distance in ssDNA has been reported to be approximately 0.4-0.6 nm.[54, 55] Applied to a $(A_{10}C_{15})_8$ sequence, transitions from A→C and C→A should occur every 4-6 nm and 6-9 nm, respectively. The $(A_{10}C_{15})_8$ sample was prepared as stated in the experimental section according to the literature.[32, 37] Thus, the phosphate groups are responsible for a tight binding to the substrate, whereas the nucleobases point upwards and can directly be accessed by the tip. In **Figures 1a** and **b**, a sketch of the experimentally realized tip-DNA arrangement is illustrated using realistic size ratios between the tip and the actual molecules.

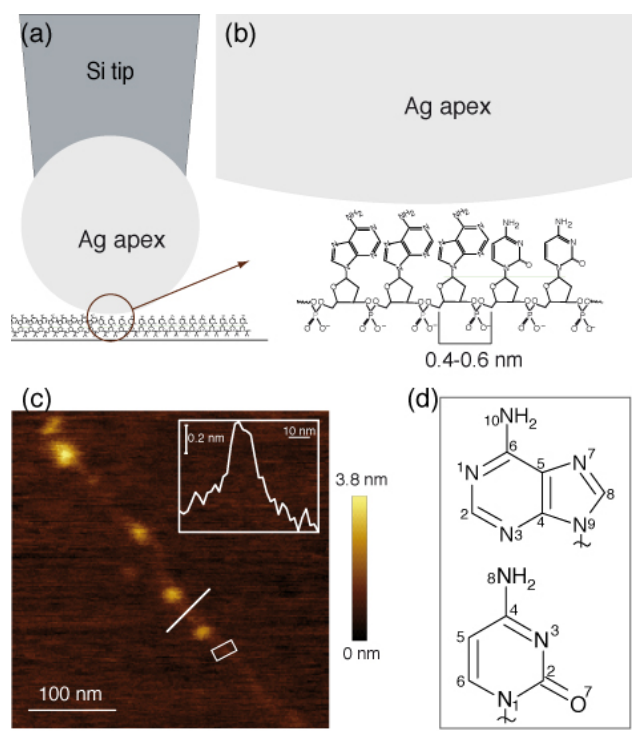

**Figure 1** (a), (b) Sketch of the tip-DNA strand arrangement on mica to demonstrate the relative sizes of the components. Tip diameter is ~15 nm. (c) AFM topography of a single $(A_{10}C_{15})_8$ strand imaged during the TERS experiment. The inset shows the cross section with the typical height of ssDNA; the rectangle indicates the region of the TERS measurement. (d) Molecular structures of adenine and cytosine with atom numbers.

Pacing along the $(A_{10}C_{15})_8$ strand with a step size of 0.3 nm, the transition from adenine to cytosine should be visible, provided that the tip offers a spatial resolution below 1 nm. In **Figure 1c**, the AFM topography imaged during the TERS experiment is depicted, and the strand height given by the cross-section (~ 0.45 nm, inset in **Figure 1c**) agrees with the reported values for ssDNA.[56] The molecular structures of the bases are given in **Figure 1d**. The measured strand width of ~ 10 nm results from a convolution between the tip and the strand. The lumps (bright spots in the topography image) originate from hair-pin loops, formed during sample adsorption, and could be easily avoided during spectral acquisition. Consecutive spectra along a line on the strand were recorded with a step size of 0.3 nm in the marked region (**Figure 1c,** rectangle). Subsequently, this line is referred to as "trace." After 25 spectra (trace: Spectra at 0-7.2 nm), the tip was shifted laterally by 1 nm, and again 25 spectra on equidistant points were recorded (retrace: Spectra at 7,2' nm-0' nm). Trace and retrace were intentionally shifted by 1 nm to pursue the effect of altered detection sites. Similar to Watanabe et al., who displaced a TERS tip on a single adenine crystal and observed significant spectral changes[44], a similar effect was expected in the present experiment. The ability of TERS to distinguish molecular sites is corroborated by recent works where the TERS tip was laterally moved (by 1 nm) on porphyrin molecules.[16, 17] In **Figure 2**, all raw spectra of the measurement are plotted. Due to the acquisition time of 5 s per spectrum, the number of points had to be limited to 50 spectra; beyond that, the accuracy of the measurement would have suffered from thermal drift effects, which were detected to be approximately 30 nm/h. Shorter acquisition times would have implied a loss of spectral information, and in the worst case, only a single band would be detected, which we considered at this stage to be insufficient for a reliable identification.

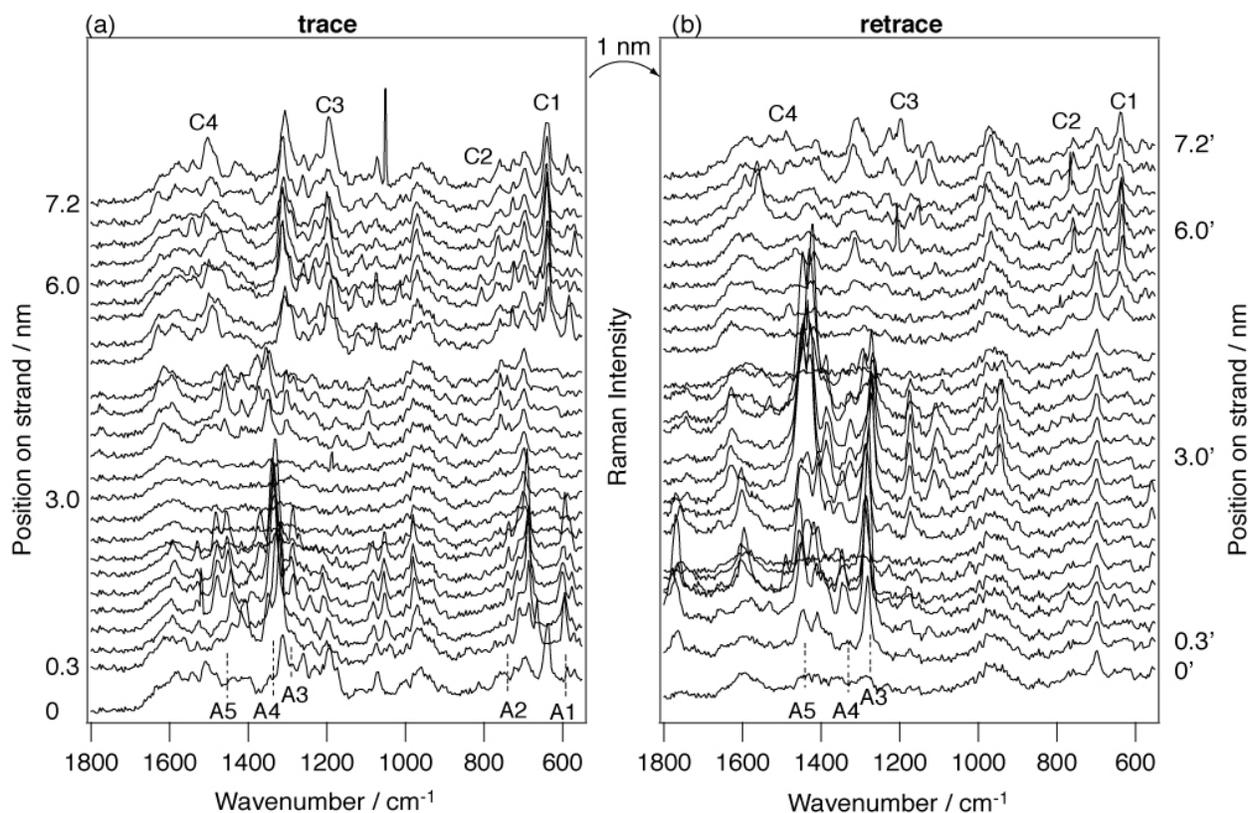

**Figure 2.** Raw TERS spectra on a $(A_{10}C_{15})_8$ single strand. (a) Trace spectra recorded on 25 consecutive points (spectra at 0-7.2 nm). (b) Retrace spectra recorded on 25 consecutive points (spectra at 7.2'-0' nm). Trace and retrace have a lateral offset of 1 nm; all spectra were recorded with an acquisition time of 5 s / spectrum; step size between every point in each trace was 0.3 nm. A1-A5 and C1-C4 indicate marker bands for adenine and cytosine, respectively.

Already at a first glance - without the need of a detailed band analysis - two spectral patterns in the trace and retrace are plainly visible. For instance, using only the ring breathing modes (marker bands A2 and C2), adenine and cytosine can unambiguously be differentiated. Surprisingly, no gradual transition with the mixed spectra (containing A and C information concurrently) is found, but instead, immediate transitions from A→C in the trace at 4.5-4.8 nm (with respect to the starting point of the measurement at 0 nm) and a transition from C→A in the trace at 0.3 nm and in the retrace at 4.8'-4.2' nm (spectrum 4.5' nm did not hold useful data) occurred. Those transitions are a strong indicator for sub-nanometer spatial resolution. The slightly varying band positions and intensities further signify that only a few molecules were locally probed.[2, 50] A loss of all signals in the spectra (trace at 2.1-3.3 nm; retrace at 4.5' nm and 1.8' nm) can be explained by a brief lost feedback. This is related to the fact that the oscillation amplitude of the tip is set to small values to optimize the turnover rate for the spectral acquisition and to keep the tip as close to the sample as possible. This, however, can lead to adverse effects on the ability to produce a stable feedback.

In addition to the characteristic ring breathing modes, four additional marker bands for adenine (non-overlapping with the cytosine bands) and three for cytosine were specified, all in accordance with previously reported band assignments (see **Table 1**; the full assignment for all bands is provided in the Supporting Information S1). It is obvious that most of the marker bands fluctuate in a broader spectral region than what has been usually reported for TERS, where band fluctuations of +/- 5 cm$^{-1}$ in single molecule experiments have been observed.[50, 51] This is because most of the bands in Table 1 are a combination of vibrational modes rather than a single mode.

**Table 1** Adenine and cytosine marker band assignment from the $(A_{10}C_{15})_8$ TERS spectra

| Wavenumber / cm$^{-1}$ | Marker band | Adenine | Marker band | Cytosine |
|---|---|---|---|---|
| 635-641 | | | C1 | $N_8$-H[33, 43, 47] |
| 689-693 | A1 | $C_6$-$N_{10}$, $C_5$-$N_7$-$C_8$, $C_4$-$C_5$-$C_6$, $N_3$-$C_4$-$N_9$[33, 57] | | |
| 735-743 | A2 | ring breath[33, 35, 36, 38, 44] | | |
| 805-811 | | | C2 | ring breath[34-37, 58] |
| 1194-1206 | | | C3 | ring def. $N_1$-$C_2$, $C_2$-$N_3$[36, 47] |
| 1275-1292 | A3 | $C_8$-H, $N_7$-$C_8$, $C_2$-$N_3$[38, 41, 44] | | |
| 1330-1349 | A4 | $C_2$-H, $C_8$-H, $N_1$-$C_6$, $C_8$-$N_9$, $N_3$-$C_4$, $C_2$-$N_3$[35, 41, 44] | | |
| 1420-1453 | A5 | $C_6$-$N_{10}$, $N_7$-$C_8$[41] | | |
| 1493-1510 | | | C4 | $NH_2$, $C_4$-$C_5$, all H[15, 35, 37, 47] |

Although the crude spectral patterns already provide ample evidence of sub-nanometer resolution, a detailed band analysis was also conducted to address spectral features that are not easily explained. To further analyze the spectra, signals were investigated by non-linear band fitting. In this procedure, only bands with a signal-to-noise-ratio (SNR) above 1.2 were considered for practical reasons. In **Figures 3a** and **b**, information of the dataset regarding band positions and intensities is summarized. Spot sizes in **Figure 3** are proportional to the signal intensity and were normalized independently to each specific band; consequently, the spot sizes can be compared only for a specific mode (e.g., between A1). An example of the fitting procedure is given in the Supporting Information S2.

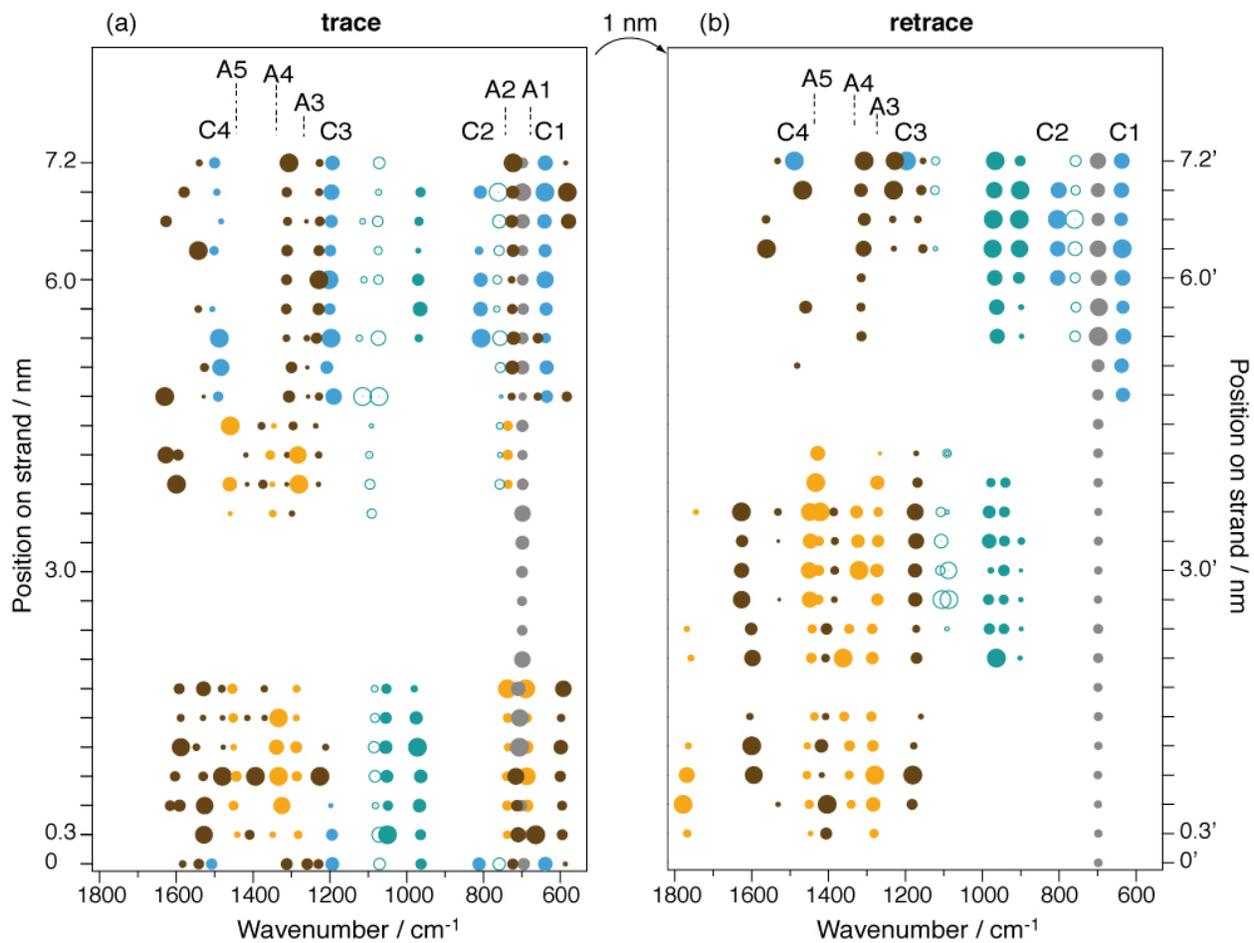

**Figure 3.** Plots of band fitting parameters vs. strand position of the spectra on (a) trace and (b) retrace of Fig. 2. Spot location corresponds to the spectral position, and spot size corresponds to the intensity. Intensities were normalized with respect to each separate band. Color code: Yellow: Adenine marker bands, blue: Cytosine marker band, brown: Bands that could be assigned to either A or C, green: Deoxyribose, dotted circle: Phosphate, grey: Mica. The assignment for adenine is indicated by A1-A5 and for cytosine by C1-C4 (see Table 1).

In addition to the marker bands of A (yellow) and C (blue), the plot shows signal contributions of mica (grey), deoxyribose (green) and phosphate (dotted circle). Band intensities of the backbone moieties are remarkably lower than those of the bases, which is in accordance with the depth resolution of TERS and is in agreement with previous TERS results on DNA.[37, 38] Signals indicated by brown spots cannot be unambiguously assigned to either adenine or cytosine and therefore were not used for distinction. To comprehend the spectral differences, a thorough interpretation of the band position and intensity variations was essential. In the following discussion the focus is on the bands of the nucleobases, and the omnipresent sugar and phosphate (backbone) bands are mentioned only if notable changes were detected.

The first spectrum in the trace has merely spectral characteristics of cytosine and is followed by a spectrum that shows only adenine signals. This indicates a transition from cytosine to adenine in one step of 0.3 nm, providing a first hint to a spatial resolution below 1 nm. For the following 1.5 nm, only adenine is detected (equivalent to six positions). A closer look at the marker bands A1-A5 reveals that the band positions and intensities vary slightly, suggesting that the consecutive adenine molecules faced the tip either in a different geometry or with different molecular sites. Of course also a combination of both effects cannot be ruled out. Again, the observations are in agreement with earlier adenine crystal studies[44], publications on Poly(A) and Poly(C) strands[37, 38] and recent observations on porphyrins.[16] Between 2.1-3.3 nm, the tip must have lost feedback, and thus, spectral information was lost. This phenomenon frequently occurs in TERS when working with low amplitudes to constantly keep the tip in a close distance to the sample and to obtain a tradeoff between TERS enhancement duty cycle and feedback stability. We decided to include this experiment in the manuscript although it is not a best-case-scenario because it reflects the reality under these experimental conditions. One always has to bear in mind that this type of experiment is very different from TERS experiments where strong resonant Raman scatterers, such as carbonnanotubes or dyes, are studied. Additionally, sample preparation is not trivial either. First, a suitable stretched and uncoiled single DNA strand has to be found with the TERS tip in AFM mode. Then, the very same tip must still be active enough to collect the TERS spectra and always ensuring that no tip contamination occurred.

Coming back to Figure 3, starting from 3.6 nm, again only adenine is detected. Because $(A_{10}C_{15})_8$ comprises ten successive adenine units, we hypothesize that the "empty" spectra likely contain masked contributions from adenine than from cytosine. Considerable spectral changes are tracked between 3.9-4.5 nm, where A1 and the sugar band at 970 cm$^{-1}$ are absent. Moreover, on approaching cytosine at 4.8 nm, the phosphate band at approximately 1100 cm$^{-1}$ splits up. Presumably, not only adenine but the entire strand had tilted from its initial geometry. Whether interactions of the electron rich aromatic systems with the electromagnetic field at the tip apex or the highly flexible chain of a ssDNA[54] caused the distortion remains unclear. From 4.8-7.2 nm, the spectra contain merely cytosine contributions as evident from the marker bands C1-C4. After a scan of 7.2 nm, the tip was shifted 1 nm sideways, and the retrace was measured. The first 9 spectra (7.2'-4.8' nm) can be assigned to cytosine, which perfectly matches the information from the trace. Interestingly, in contrast to the trace data, not all cytosine marker bands are detected. A band at 900 cm$^{-1}$ can be assigned to the carbohydrate backbone. Both facts together most likely indicate a changed tip-strand orientation as noted above for adenine along the trace. Due to the "empty" spectrum at 4.5' nm, the C→A transition can be estimated to occur only between 4.8'-4.2' nm. Nevertheless, starting from 4.2' nm, all spectra can be exclusively assigned to adenine. Surprisingly, neither A1 nor A2 (ring breathing mode) is identified until the end of the measurement. The absence of the ring breathing mode was unexpected because it usually is a prominent band. However, it can be explained by the flat orientation of the entire purine ring with respect to the tip. In such a geometry, the in-plane mode A2 does not experience an enhancement by the confined field at the tip apex, which is known to be "active" mainly in the z-direction and therefore preferably observed for tilted molecules.[27] Similar observations have been reported on aromatic amino acid monolayers when adsorbed with the phenyl ring lying flat on the substrate[59] To further verify this assumption, theoretical calculations were pursued. To that end, semi-empirical calculations on a small ssDNA model consisting of one cytosine and three adenine bases have been performed (for details see Supporting Information S3). We scanned the energy profile of the torsion of adenine from the up-pointing geometry to being laid flat on the mica surface. The calculations show that the energy barrier for such torsion is merely 0.36 eV, and the energy level of the new geometry is only 0.17 eV higher. Because this is not a relaxed scan, the real barriers are expected to be even lower. Based on the calculations, we assume that the adenine orientation with respect to the sample surface must have changed from an upright geometry to an arrangement with the ring system parallel to the surface. Evidently all molecules within the 3.9 nm adenine region had adopted the same geometry. Further evidence for a modified orientation is the presence of additional bands at approximately 1750 cm$^{-1}$ and 1778 cm$^{-1}$, which is an unusual spectral region for adenine. Tentatively those bands can be assigned to $NH_2$ and pyrrole ring modes, which is again supported by the theoretical calculations (see Supporting Information S4, S5). Band intensities continuously decrease throughout the last spectra (1.5'-0' nm), and neither deoxyribose nor phosphate signals were detected. This was most likely due to thermal drifts of the entire setup until the tip was finally off the strand.

The spectral changes related to cytosine are not as prominent as for adenine, which also indicate a change of the pyrimidine ring orientation from trace to retrace. On both lines, sharp base transitions are detected at 0-0.3 nm, 4.5-4.8 nm and between 4.8'-4.2' nm. Similar to adenine, alterations of the cytosine marker band intensities and positions can be associated with not only the lateral offset of the tip from trace to retrace but with changes of strand geometry and interacting molecular moieties. The marker band C1 is detected in all cytosine spectra of trace (5.4-6.9 nm) and retrace (7.2'-4.8' nm), whereas C3 and C4 are not continuously present. The same is valid for the ring breathing mode C2 that is detected only sometimes. Furthermore, the sugar band at 900 cm$^{-1}$ appears again on the retrace, whereas the phosphate band at approximately 1120 cm$^{-1}$ can be only partly detected. These observations all suggest once more a molecular reorientation of the bases in that the pyrimidine ring was oriented parallel towards the surface.

In summary, our experiment on the $(A_{10}C_{15})_8$ strand clearly demonstrates a reproducible sub-nanometer resolution of AFM-TERS at ambient conditions. A value of less than 0.5 nm (from the sharp A→C or C→A transitions, the resolution is 0.3 nm) could be estimated from the precise base-to-base transitions. Interestingly, this correlates with the possible step size of the AFM employed. Along the measured lines, a set of adenine spectra is followed by a set of cytosine spectra and vice versa. A smooth transition from C to A, as one would expect for a predicted resolution of approximately 3-4 nm for our tips and the selected sample, is clearly not observed. Further information was gained on the molecular conformations, evident from the presence and absence of distinct bands in the 1 nm relocated probed lines. These promising results raised the question of whether single embedded adenine moieties can be discerned from surrounding cytosines. For this experiment, ssDNA with single adenine embedded in varying numbers of cytosine bases was used. Specifically, the sequence $(ACAC_3AC_5AC_{10}AC)_8$ was prepared following the procedure for $(A_{10}C_{15})_8$. Here, a single line with 50x0.3 nm equidistant acquisition points along the strand was measured before the tip became contaminated. In **Figure 4a**, four selected spectra (region 3.3-4.2 nm) are shown with highlighted marker bands that are in accordance with that previously specified for $(A_{10}C_{15})_8$. Again, transitions from C→A→C are clearly visible. In all spectra, the ring breathing modes (A2, C2) are omnipresent and are accentuated by zooming into the respective regions in **Figure 4b** (the complete spectral dataset is provided in the Supporting Information S6 and the full band assignment in S1). All spectra were assessed as described in the first experiment, and the fitting parameters of the marker bands are plotted in **Figure 4c**. Again, the spot size is proportional to the signal intensity and is normalized to each specific band independently, meaning that the spot sizes of different modes cannot be compared directly. A plot of the fitting parameters of all marker bands does not provide additional information but can nevertheless be found in the Supporting Information S7.

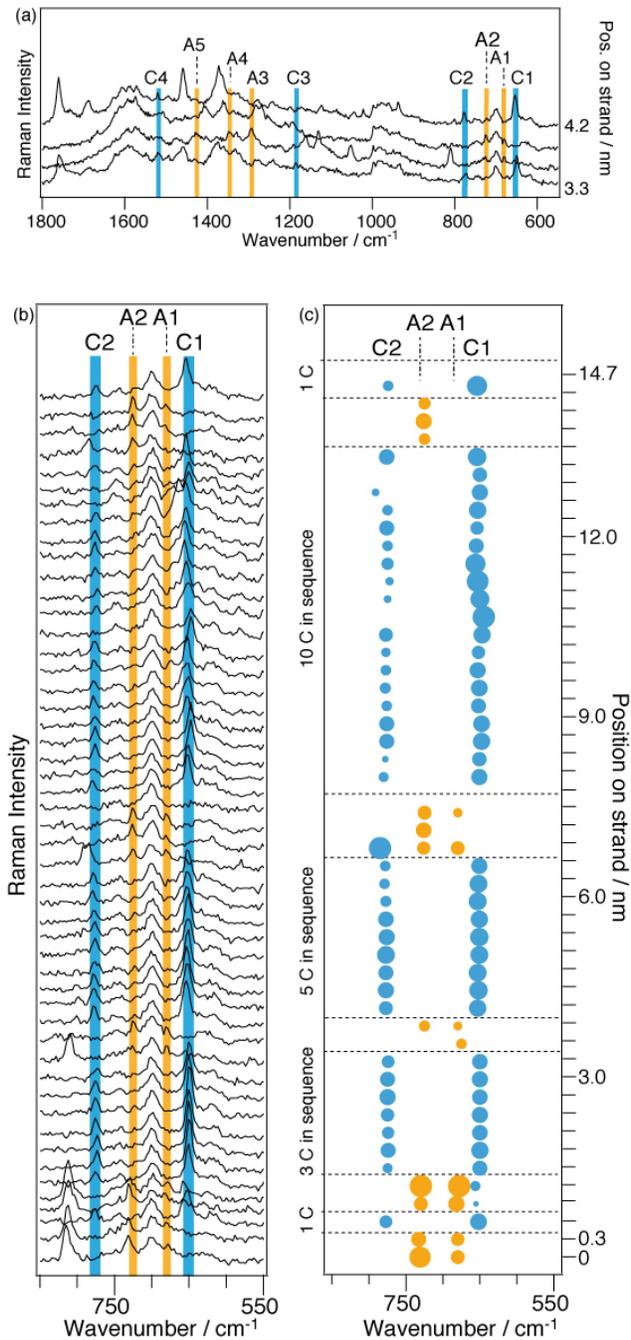

**Figure 4** (a) Four selected raw spectra starting at position 3.3 nm of the TERS experiment on a (ACAC$_3$AC$_5$AC$_{10}$ AC)$_8$ single strand. Step width: 0.3 nm. A1-A5 indicate marker bands for adenine, and C1-C4 indicate marker bands for cytosine. (b) All 50 untreated spectra collected on the strand within the spectral region show the marker bands of adenine (A1, A2) and cytosine (C1, C2). (c) Plot of the fitting parameter band position and intensity vs. the strand position of the spectra in (b). Spot location corresponds to the spectral position, and the spot size corresponds to the intensity. Intensities were normalized with respect to each separate band; adenine - yellow, cytosine - blue.

From **Figure 4**, is it evident that the spectra of the respective bases do not change as substantially as in the previous experiment. Again, clear transitions from A→C and C→A are detected repetitively, confirming the sub-nanometer resolution found in the previous experiment. It is worth to note that the experiments on the different strands were performed independently on different days with different tips. Although the experimental conditions (temperature, humidity) varied slightly, the similarity of the results validated the reproducibility of the TERS measurements. Interestingly, each embedded adenine contributes not only to a single spectrum but to two and sometimes even three consecutive spectra, as observed from positions 0-0.3 nm, 0.9-1.2 nm, and 3.6-3.9 nm, etc. The same observation is made for cytosine, where one C yielded one spectrum at 0.6 nm, but three consecutive moieties provided seven spectra, and five C in a row resulted in 10 spectra, etc. It should also be mentioned that the coincidence of the starting point of the measurement and the beginning of the sequence was by pure chance. As already stated above, base-to-base distances in ssDNA are known to be flexible (0.4-0.6 nm) and most likely were larger in $(ACAC_3AC_5AC_{10}AC)_8$ than in $(A_{10}C_{15})_8$, leading to the detection of each base in numerous consecutive spectra. Nevertheless, from the spectra, an increasing number of cytosine moieties in the strand, separated by single adenine molecules, can be deduced. It is evident that in the second experiment, the ring breathing modes are significantly shifted to lower wavenumbers, i.e., A2 from 740 cm$^{-1}$ to 726-730 cm$^{-1}$, and C2 from 809-805 cm$^{-1}$ to 776-784

cm$^{-1}$. All other marker bands are also shifted, albeit not that strongly. These band shifts indicate a different chemical environment of the bases in the examined strand area compared to the previous sample. Even more obvious is the detection of the strongly enhanced new bands for cytosine at 1362-1383 cm$^{-1}$, 1689 cm$^{-1}$, and 1738-1755 cm$^{-1}$. The latter bands have been reported in UV resonance Raman spectroscopy experiments of the protonated imino-tautomer of cytosine.[48] We assume that it was the slightly modified sample preparation that rendered the iminol-tautomer more stable than the commonly present amino-keto form, but this circumstance does not affect the results of the TERS experiment. To demonstrate the reproducibility of the presented experiments, measurements were repeated with different tips. The raw spectra and fitted data obtained from a freshly prepared (ACAC$_3$AC$_5$AC$_{10}$AC)$_8$ sample is provided in the Supplementary Information (S8 and S9). Again, the sequence could be deduced from consecutive spectra recorded along the main axis of the strand. In a fourth experiment (see S10), a trace-retrace measurement along a cytosine region of (ACAC$_3$AC$_5$AC$_{10}$AC)$_8$ was measured similar to the very first experiment in this paper. All presented and discussed observations were confirmed in these experiments. In particular, it is evident that even a small lateral offset of the two traces results in slightly different spectral appearances, in line with the slightly different chemical environments if the "touching point" is shifted even by merely a bond length.

All measurements unambiguously illustrate that it is possible to detect single adenine molecules embedded in varying numbers of consecutive cystosine units in ssDNA with a spatial resolution below 1 nm.

**Conclusion**

In the present contribution, TERS under ambient conditions was used for the detection of base-to-base transitions in single-stranded DNA. Strands composed of adenine (A) and cytosine (C) were probed along lines on 0.3 nm equidistant acquisition points. On the trace and retrace of $(A_{10}C_{15})_8$, adenine-to-cytosine transitions were detected directly, demonstrating sub-nanometer spatial resolution. The transitions and also spectral variations indicated that single bases under the tip contributed to each spectrum and revealed the high structural sensitivity of the experiment. This was confirmed in three individual experiments on $(ACAC_3AC_5AC_{10}AC)_8$ single strands. Here, adenine molecules embedded in sets of cytosine could repeatedly be identified and distinguished alongside with the precise A→C and C→A transitions. All experiments plainly illustrated the sensitivity of TERS to detect molecular geometry changes with extremely high spatial resolution. The presented results are very promising for future projects where the identification and localization of substituted bases in DNA strands will be targeted.

ASSOCIATED CONTENT

**Supporting Information**.

S1: Band assignment of TERS spectra recorded on DNA single strands.

S2: Exemplified non-linear Lorentzian band-fitting of TERS spectrum at 5.4 nm on $(A_{10}C_{15})_8$.

S3: Quantum chemical calculations of structural alterations of ssDNA.

S4: Adenine structure with marked positions of the silver atom for spectra calculation.

S5: Calculated Raman spectra of adenine with a silver atom at different positions.

S6: Complete raw spectral dataset of the TERS measurement on $(ACAC_3AC_5AC_{10}AC)_8$.

S7: Plot of all fitted marker bands identified in the TERS spectra of $(ACAC_3AC_5AC_{10}AC)_8$.

S8: Plot of all fitted marker bands identified in the reproducibility TERS experiment of a second $(ACAC_3AC_5AC_{10}AC)_8$ strand.

S9: Complete raw spectral dataset of the reproducibility TERS measurement on a second $(ACAC_3AC_5AC_{10}AC)_8$ strand.

S10: Raw spectra and plot of fitted marker bands of the reproducibility trace and retrace experiments on a second $(ACAC_3AC_5AC_{10}AC)_8$ strand.

This material is available free of charge via the Internet at http://pubs.acs.org.

AUTHOR INFORMATION

**Corresponding Authors**

*Email: volker.deckert@uni-jena.de, tanja.deckert-gaudig@leibniz-jena.de

**Author Contributions**

The manuscript was written through contributions of all authors. All authors have given approval to the final version of the manuscript. X.-M. Lin performed TERS experiments on $(A_{10}C_{15})_8$. X.-M. Lin and T. Deckert-Gaudig analyzed the corresponding data. P. Singh and T. Deckert-Gaudig performed the TERS experiments on $(ACAC_3AC_5AC_{10})_8$. P. Singh and T. Deckert-Gaudig analyzed the corresponding data. M. Siegmann and S. Kupfer performed calculations on $(A_{10}C_{15})_8$, and Z. Zhang on the cytosine conformation. S. Gräfe was responsible for the theoretical investigation. T. Deckert-Gaudig and V. Deckert prepared the manuscript. V. Deckert was responsible for the experimental implementation, providing conceptual advice and supervising the project. All authors contributed extensively to the work presented in this paper.

**Notes**

The authors declare no competing financial interest.

ACKNOWLEDGMENT

We gratefully acknowledge financial support by the German science foundation (DFG, DEP4TERS, FR 1348/19-1). ZZ acknowledges financial support from the Alexander von Humboldt foundation. All calculations have been performed at the Universitätsrechenzentrum and with HP computers at the Theoretical Chemistry department of the Friedrich-Schiller-University Jena.